\documentstyle[aaspp]{article}

\def\stacksymbols #1#2#3#4{\def\theguybelow{#2}
	\def\verticalposition{\lower#3pt}
	\def\spacingwithinsymbol{\baselineskip0pt\lineskip#4pt}
	\mathrel{\mathpalette\intermediary#1}}
\def\intermediary #1#2{\verticalposition\vbox{\spacingwithinsymbol
	\everycr={}\tabskip0pt
	\halign{$\mathsurround0pt#1\hfil##\hfil$\crcr#2\crcr
		\theguybelow\crcr}}}
\def\lta{\stacksymbols{<}{\sim}{2.5}{.2}}
\def\gta{\stacksymbols{>}{\sim}{3}{.5}}

\begin{document}
\title{SELF-GENERATED MAGNETIC FIELDS IN GALACTIC COOLING FLOWS$^1$}

\author{William G. Mathews$^2$ and Fabrizio Brighenti$^{3}$}

\affil{$^2$University of California Observatories/Lick Observatory,
Board of Studies in Astronomy and Astrophysics,
University of California, Santa Cruz, CA 95064\\
mathews@lick.ucsc.edu}

\affil{$^3$Dipartimento di Astronomia,
Universit\`a di Bologna,
via Zamboni 33,
Bologna 40126, Italy\\
brighenti@astbo3.bo.astro.it}






\begin{abstract}

Interstellar magnetic fields in elliptical galaxies have their origin
in stellar fields that accompany normal mass loss from an evolving
population of old stars.  It is likely that these seed fields are
amplified by interstellar turbulence which in turn is driven by
stellar mass loss and supernova events.  Since the local turbulent
velocity is likely to exceed the global velocity of the interstellar
cooling flow, magnetic fields are expected to be disordered and this
is indicated by numerous observations.  Further amplification occurs
as these tangled fields are compressed in the inward moving cooling
flow of the interstellar gas.  Near the centers of galactic cooling
flows, where the gas radiates away most of its thermal and
gravitational energy, magnetic stresses are expected to dominate. We
study here the time-dependent growth of interstellar magnetic fields
in elliptical galaxies and demonstrate that fields similar in strength
to those observed can be generated solely from natural galactic
processes.

Although interstellar turbulent velocities and therefore the
equipartition field $B_e$ are larger near the galactic center, the
interstellar field throughout elliptical galaxies can be determined by
the outermost turbulent regions in the interstellar gas.  This occurs
whenever the increase in the field strength due to compression in the
galactic cooling flow exceeds the rate of field amplification by local
turbulence, i.e. when $B > B_e$. The magnitude of the interstellar
field depends on the most distant radius in the galactic ISM at which
the turbulent dynamo process can successfully have amplified stellar
seed fields.  Because of the long hydrodynamic flow times in galactic
cooling flows, currently observed magnetic fields may result from
periods of turbulent field amplification that occurred in the outer
galaxy in the distant past.  In our most optimistic turbulent dynamo
models in which field reconnection is ignored, tangled
interstellar magnetic fields of $B
\sim 1-10~(r/10~{\rm kpc})^{-1.2}\mu$G are typical;
these are consistent with observed fields.  However, the sensitivity
of the galactic magnetic field to poorly known historical turbulent
conditions in the outer galaxy complicates estimates of field
strengths that can be produced internally.  Internal fields in
ellipticals may also result from ancient galactic mergers or from
shear turbulence introduced at the boundary between the interstellar
gas and ambient cluster gas.

\end{abstract}

\newpage

\section{INTRODUCTION}

It is likely that gas observed in the central regions of 
elliptical galaxies is dominated by magnetic stresses.
The chaotic, non-equilibrium  appearance of dusty clouds of cool gas 
apparent in HST observations in the cores of many elliptical
galaxies (e.g. van Dokkum \& Franx 1995)
is often ascribed to the effects of a recent merger. 
However, it may be more likely that 
the disorganized appearance of these clouds
reflects the well known inability of magnetic
fields to reach quiescent configurations in differentially 
rotating, gravitationally bound fluids.
In addition it is plausible that strong magnetic forces
in the gas that cools near the centers of elliptical galaxies 
may influence the initial mass function or other aspects of
star formation associated with galactic cooling flows.

The interstellar medium in ellipticals and 
its magnetic field can be regarded as a result of 
cumulative stellar mass loss from an evolving population
of old stars.
Gas ejected from orbiting stars thermalizes 
a significant fraction of its kinetic
energy and settles into a quasi-hydrostatic equilibrium 
with temperatures near the virial temperature of the 
galactic potential, $T_{vir} \sim 10^7$ K.
The hot gas then loses energy by emitting the thermal X-rays
that are observed and slowly sinks in the galactic potential
where it is compressed and maintained approximately isothermal.
Although these subsonic inflows are commonly referred to as 
galactic ``cooling flows,'' the cooling occurs only 
in the central regions where radiative losses finally overwhelm
compressional heating.

The hot interstellar medium in elliptical galaxies operates like a
magnetic distillery, first amplifying fields introduced in 
stellar ejecta then concentrating this magnetic energy in 
high density gas near the center of the galactic 
cooling flow where intense radiative cooling occurs.
Small fields contained in winds and planetary nebulae 
expelled from red giants are thought to be 
amplified further by turbulent motions in the 
interstellar medium (ISM).
Even in the presence of turbulent amplification
magnetic stresses at large galactic radii are expected 
to be small compared to the pressure in the hot 
interstellar gas.
However, the interstellar gas compresses as it slowly flows 
toward the galactic center and the field strength 
must also grow by flux 
conservation $B \sim \rho^{2/3}$.
The ratio of magnetic to (isothermal) gas
pressure also increases with gas density, 
$B^2/\rho \sim \rho^{1/3}$.
Finally, as gas cools near the central parts of the
cooling flow, the thermal support of the gas drops precipitously 
and magnetic stresses are distilled out to support the 
remaining ionized 
gas against gravity and external gas pressure.

Soker \& Sarazin (1990) discussed the spherical,
steady state evolution of 
completely disordered, frozen-in 
fields in cluster-scale cooling flows.
They chose the field strength to be 1 $\mu$G at a radius
of 50 kpc.
Within $\sim 5$ kpc in their models 
the magnetic pressure grew by 
compression in the cooling flow 
to be comparable to the gas pressure, 
but further amplification was not thought to be possible
because of loss of field by reconnection.
Since the cooling flow accelerates as the 
galactic center is approached, Soker \& Sarazin noted
that an initially random field would develop a large
scale radial component.
More recently Christodoulou \& Sarazin (1996) computed
the time-dependent evolution of large scale toroidal
and axial fields in cluster cooling flows adapted to
cylindrical geometry.
Although $B^2/8 \pi$ may be 
small relative to the local gas pressure, the
origin of $\sim 1 \mu$G fields assumed 
by Soker \& Sarazin and Christodoulou \& Sarazin 
at the start of their calculations must be understood. 
Primordial field production prior to galaxy formation 
results in much smaller 
fields ($\lta 10^{-15}$ G) (Rees 1987; Lesch \& Chiba 1995).
But the high metallicity in gas in clusters of galaxies 
implies that much of this gas was enriched and 
expelled by Type II supernovae out of galaxies in which
interstellar field growth mechanisms were already available.
Thus cluster-scale magnetic fields probably have their
origin in fields generated in the ISM of cluster 
galaxies.

Our interest here is to explore the growth of interstellar 
magnetic fields in individual elliptical galaxies in a 
self-consistent fashion beginning with small seed
fields contained in stellar ejecta.
We assume that stellar dynamos are efficient
in producing fields in stellar envelopes. 
Stellar mass loss eventually transports these fields
into the ISM where the field strength is greatly diminished
by adiabatic expansion.
However, recent theoretical work 
(Lesch \& Bender 1990; Moss \& Shukurov 1996)
has suggested that some the kinetic 
and thermal energy that 
accompanies stellar mass loss processes 
in ellipticals can also drive
turbulent motions in the ISM which in turn amplify the magnetic
field to $\sim 1 - 10~\mu$G.
In this paper we shall assume with these authors that
relatively small interstellar magnetic fields can grow by
this turbulent dynamo.
We describe how large scale 
fields may develop from random fields,
but we suppose for simplicity that most of the field remains
random on small spatial scales ($\lta 0.1$ kpc say).
The growth of magnetic forces in the ISM 
may ultimately be limited by field reconnection
which we allow for in an approximate way.

Unfortunately, the physical processes that control the 
turbulent growth of galactic magnetic fields are poorly 
understood and controversial.
In our Galaxy the mean field
$\alpha - \Omega$ dynamo has been used to describe the
amplification  of 
small seed fields to equipartition
with interstellar turbulence 
(Ruzmaikin, Shukurov, \& Sokoloff 1988).
Equipartition of magnetic and turbulent energy densities
is in fact observed.
Fields in spiral disks can grow both by turbulence
and differential rotation.
Enhancement of very small seed fields ($\ll 10^{-6}$ G) 
by the turbulent dynamo process is attractive because
amplification by differential rotation in the Galactic disk 
is limited by the relatively few rotations (50 - 100 times)
experienced by the Galaxy during its lifetime.
However, several authors (Vainshtein \& Cattaneo 1992; 
Kulsrud \& Anderson 1992) have argued that the linear 
turbulent dynamo cannot amplify very small seed fields 
to equipartition values.
Kulsrud \& Anderson assert, for example, that the field strength
can grow exponentially 
with time to equipartition only on small spatial scales
where it is dissipated into heat while the mean field
at intermediate and large spatial scales 
grows to saturation far below equipartition values.
The (damping) reaction of the field on the background
hydrodynamic turbulence is also considered.
These authors claim that the failure of
dynamo theory to account for fields of equipartition 
strength implies that Galactic fields must 
be primordial.
But fields expected in proto-Galactic gas 
are likely to be much lower than those currently
observed (Lesch \& Chiba 1995; Beck et al. 1996).
Field (1995) has reviewed this apparent difficulty with the mean
field dynamo theory and suggests that a more appropriate 
application of
the Kulsrud-Anderson formalism would be 
to study the {\it steady-state} distribution 
of spectral field components in response to a constant
input of turbulent energy on large scales. 
But an extension of the theory to include non-linear
terms may also be necessary.
In their recent comprehensive review of galactic 
fields Beck et al. (1996) note the apparent inadequacy of 
the linear theory in addressing turbulent growth;
a non-linear turbulent dynamo theory 
would include inverse cascade of small
scale field energy to larger scales as well as non-linear
aspects of small scale dissipation.
Finally, Beck et al. (1996) note that 
fields observed in the Galactic ISM are not
concentrated to small spatial scales as the linear 
turbulent dynamo theory would predict.

In view of these difficulties and possible shortcomings
of mean field dynamo theory, we adopt here a simplified model
that allows turbulent growth of mean interstellar
fields in elliptical galaxies.
We are encouraged to take 
this deliberately optimistic approach by analogy with 
equipartition fields observed 
in the ISM of our own Galaxy
and by the arguments of Moss \& Shukurov (1996)
and others
that turbulent dynamos are likely to be effective 
in elliptical galaxies.
By assuming the validity of turbulent dynamos, we 
can estimate the (maximum) field strengths expected
in the ISM of elliptical galaxies from internal processes
alone; no {\it ad hoc} initial galactic field is required.

We have been motivated to study the evolution 
of magnetic fields in ellipticals 
by our recent work on the nature of cooling flows in 
slowly rotating ellipticals (Brighenti \& Mathews 1996).
We showed there that a small rotation of the stellar component,
similar to those observed 
in most large ellipticals ($\sim 50 - 100$ km s$^{-1}$),
causes the interstellar gas to cool onto a very large
disk comparable in size to the half-light radius.
Most of the new gas reaching the disk arrives at its outer
edge so that the cooling disk slowly grows with the age
of the galaxy.
As hot gas approaches the relatively high density regions
near the outer edge of the cooling disk, its thermal X-ray
image resembles a bright disk or torus in the equatorial plane.
Near the equatorial plane the gas spins up 
to the circular velocity which is much larger
($\sim 400 - 500$ km s$^{-1}$) than that of the stellar system.
We estimated that a small, dynamically insignificant magnetic
field would amplify by a factor of $\gta 5$ in the differentially
rotating flow as hot gas approaches the edge of the cooling 
disk.
We speculated that the corresponding increase in magnetic forces,
by $\gta 25$, might be sufficient to limit the maximum compression
and X-ray visibility of the hot torus just beyond the edge
of the cooling disk.
If interstellar magnetic fields are sufficiently large,
the X-ray images of rotating ellipticals
may appear to be more circular than those predicted by 
Brighenti \& Mathews (1966).
This type of magnetic influence
on the X-ray image may be necessary to account for the
apparent circularity of {\it Einstein} and ROSAT images
of several bright ellipticals in the Virgo cluster.
Alternatively, the less flattened X-ray appearance of these galaxies 
could simply result from a fortuitous orientation 
of galactic rotation axes along the line of sight.
In any case, the growth and evolution of magnetic fields in 
spherical, non-rotating galaxies
which we discuss here is a necessary first step
toward a more complete study of field
amplification and X-ray morphology in slowly rotating ellipticals.

Our approach to field amplification 
is similar to that of Moss and Shukurov (1996)
who describe the growth of magnetic fields in
the turbulent interstellar 
environment energized by supernova explosions and
the orbital energy of stellar ejecta.
We extend the discussion of Moss and Shukurov 
to consider how the field is further increased 
(i) by compression as the cooling flow approaches the galactic core
and (ii) by the greater interstellar 
turbulence expected in past times 
when the stellar mass loss and supernova rates were both larger.
Our calculation is necessarily approximate 
since many theoretical
aspects of the turbulent dynamo process and field reconnection are 
based on dimensional arguments.
Our models are exploratory, not definitive.

We begin with a brief review of magnetic fields observed in elliptical
galaxies then proceed to the description of our time-dependent
models and the resulting magnetic field distributions.

\section{OBSERVATIONAL EVIDENCE FOR MAGNETIC FIELDS IN ELLIPTICALS}

The common signposts for the presence of magnetic fields 
in spiral galaxies -- 
polarization of aligned dust grains, polarized emission
from relativistic electrons and Zeeman splitting -- 
are not available in ellipticals 
due in part to the relative absence of dust 
(Tsai \& Mathews 1996a; b),
the infrequency of supernova explosions and the absence of
distributed cold HI gas.
By far the best means of detecting fields in ellipticals
is through the differential Faraday depolarization of radio sources
by fields within the hot, X-ray emitting 
cooling flow gas which lies along the line of sight.
In most FR II-type double radio sources the jet that powers
the radio lobes is visible at radio frequencies on 
only one side.
Laing (1988) first noted that the radio 
polarization was systematically
higher on the jet-side which is also the near side, assuming that 
radiation
from the jet is made more visible by relativistic Doppler beaming.
If near and far 
lobes are otherwise identical, the depolarization of the
more distant lobe must arise in the galactic ISM, not within
the lobes or in a shell of gas adjacent to the radio lobe. 
The body of evidence supporting this correlation
has grown considerably, both for lobes in large ellipticals
and in small clusters of galaxies 
(Garrington et al. 1988; Garrington,
Conway, \& Leahy 1991; Garrington \& Conway 1991).
The depolarization asymmetry is particularly strong for
projected lobe separations of galactic dimension, $\lta 100$ kpc.
By comparing the relative 
Faraday depolarization of two lobes it is
possible to determine the mean product of the electron density
and net line of sight field strength to the center of the
galaxy.
A typical value is 
$\langle n B_{\parallel} \rangle \approx 
10^{-3}$ cm$^{-3}$ $\mu$G, implying $B \sim 1 - 10$ $\mu$G
for plasma densities characteristic of the hot interstellar gas.
These values are based on models in which the magnetic
field is uniform in many small cells which are randomly 
oriented along the line of sight through the hot ISM (Burn 1966).
Randomly directed fields are also implied by 
the patterns of differential radiofrequency
Faraday rotation observed at high spatial resolution 
across the face of extended radio lobes or regions 
(e.g. Strom \& J\"{a}gers 1988; Owen, Eilek, \& Keel, 1990;
Clarke, Burns, \& Norman 1992).
Magnetic fields even larger than those indicated above may
be present if they are tangled on spatial scales much less than
the radio resolution limit.
Finally, 
Greenfield, Roberts, \& Burke (1985) observed differential radio
polarization in two gravitationally-lensed 
images of a distant quasar which they attribute to magnetic
fields in the hot ISM of the intervening CD galaxy lens.

We conclude that the hot gas in elliptical galaxies 
that participates in galactic scale cooling flows contains 
magnetic fields that
are comparable to those in spirals and
that a significant component of this field is spatially 
disordered.


\section{MAGNETIC COOLING FLOWS}

The equations that describe the evolution of hydromagnetic
cooling flows in elliptical galaxies are the usual conservation
equations modified with appropriate source terms:
$${\partial \rho \over \partial t}
+ {\bf \nabla}\cdot \rho {\bf v}
= \alpha \rho_*,\eqno(3.1)$$
$$\rho \left[
{\partial {\bf v} \over \partial t}
+ ({\bf v} \cdot {\bf \nabla}){\bf v} \right]
= -{\bf \nabla}P 
+ {1 \over 4 \pi} ({\bf \nabla \times B}){\bf \times B}
- \rho {\bf \nabla} \Phi
- \alpha \rho_* ({\bf v} - {\bf v}_*),\eqno(3.2)$$
and
$$\rho \left[
{\partial \varepsilon  \over \partial t}
+ ({\bf v} \cdot {\bf \nabla}) \varepsilon \right]
= {P \over \rho} \left[
{\partial \rho  \over \partial t}
+ ({\bf v} \cdot {\bf \nabla}) \rho \right]
- { \rho^2 f L(T) \over m_p^2}$$
$$+ \alpha \rho_* \left[ \varepsilon_o - {P \over \rho}
- \varepsilon + {1 \over 2}
| {\bf v} - {\bf v}_* |^2 \right] 
+ {B^2 \over 8 \pi \tau_{rec}}.\eqno(3.3)$$
The galactic potential $\Phi$ is appropriate for a stellar
density distribution $\rho_*(r)$ and a dark matter halo.
Except for the magnetic force term in the equation of motion
(3.2) and heating due to reconnection of magnetic fields,
the last term in equation (3.3), 
these equations are identical to those used by
Brighenti and Mathews (1996).

In the absence of plasma resistivity and field sources,
the magnetic field evolves according to the usual frozen-in 
condition:
$${\partial {\bf B} \over \partial t} =
{\bf \nabla \times}({\bf v \times B}).\eqno(3.4)$$
However, for applications to the interstellar medium of
elliptical galaxies, it is necessary to include source terms due to 
fields ejected by evolving stars, turbulent growth, 
and the loss of field due to reconnection. 
These modifications will be developed in the following sections.

Although the non-magnetic source terms in 
equations (3.1) - (3.3) have been described in detail by
Brighenti and Mathews (1996), for completeness we shall 
briefly review them again here.
New gas is introduced into the interstellar medium
by mass ejection from normally evolving stars
formed in a 
single burst at time $t = 0$.
The specific rate of mass loss is proportional to the 
local stellar density
$\alpha_* \rho_*$ (gm cm$^{-3}$ s$^{-1}$) where
$$\alpha_*(t) = \alpha(t_n) (t/t_n)^{-1.3}\eqno(3.5)$$
(Mathews 1989).
Here $t_n = 15$ Gyr represents the present time
and $\alpha(t_n) = 5.4 \times 10^{-20}$ s$^{-1}$,
valid for a wide variety of initial mass functions
(Mathews 1989).
On average this new gas appears in the galaxy at zero velocity
relative to the net local stellar velocity ${\bf v}_*$
which is non-zero only for rotating galaxies.
The negative source term in equation (3.2) represents the
momentum drag on the local gas flow ${\bf v}$ as new gas
created with velocity ${\bf v}_*$ is accelerated to 
the local flow velocity ${\bf v}$.

Equation (3.3) for the conservation of specific thermal energy
$\varepsilon = 3 k_B T / 2 \mu m_p$ ($m_p$ is the proton mass)
contains a negative term representing optically thin
radiative losses by thermal emission as
computed by Raymond, Cox \& Smith (1976):
$n_e n_H L(T)/\rho = f \rho L(T) /m_p^2$ ergs s$^{-1}$
gm$^{-1}$ where $f$ is related to the mean molecular
weight by $f = (2 + \mu)(4 - 3\mu)/25 \mu^2$;
we shall assume $\mu = 0.5$.
Additional source terms in equation (3.3) 
describe the heating of the
gas by the thermal and kinetic energy of stellar ejecta,
including supernova explosions.
After new gas enters the flow with mean velocity ${\bf v}_*$
it is accelerated to
the local gas velocity, dissipating energy
at a rate $ \alpha_* \rho_*
| {\bf v} - {\bf v}_* |^2 / 2$ per unit volume.
The dissipational heating resulting from random stellar motions and
supernova explosions is represented by a mass-weighted
characteristic temperature for the new gas,
$$T_o = (\alpha_* T_* + \alpha_{sn} T_{sn})/\alpha
\eqno(3.6)$$
where $\alpha = \alpha_* + \alpha_{sn} \approx \alpha_{*}$ 
and $\varepsilon_o = 3 k_B T_o / 2 \mu m_p$
is the corresponding specific thermal energy.
The term $P/\rho$ in the source term of equation (3.3)
represents the
work done by the stellar ejecta on the local ambient
gas as it expands from an initially dense cloud toward
pressure equilibrium with the interstellar gas.

The solar or subsolar iron abundance observed in the
interstellar medium of ellipticals indicates that
the (Type Ia) supernova rate is low and that it has
been low in recent Gyrs (Loewenstein and Mathews 1991;
Loewenstein et al. 1994).
Observations of the current supernova rate 
in elliptical galaxies also suggest rather low values,
$\sim 0.1$ SNu 
(van den Bergh, McClure, \& Evans 1987;
Cappelaro et al. 1993)
where 1 SNu corresponds to one supernova
per $10^{10}$ $L_B$ every 100 years.
The past rate of supernova explosions is even more uncertain,
but it is generally assumed that the rate was 
higher when the galaxy was younger.
We shall assume a power law time dependence of the form
$$\alpha_{sn} T_{sn} =
2.13 \times 10^{-8}~ {\rm SNu}~ (E_{sn}/10^{51} {\rm ergs})
h^{-1.7} L_B^{-0.35} (t / t_n)^{-1/2} ~~~ {\rm Ks}^{-1}
\eqno(3.7)$$
where $L_B$ is in units of the solar value
$L_{\odot B} = 4.98 \times 10^{32}$ ergs s$^{-1}$
and the reduced Hubble constant is assumed to be 
$h = H/100  = 0.75$.
The SNu parameter defined by 
equation (3.7) cannot be compared directly
with observed values since other 
uncertain parameters (such as $E_{sn}$)
also appear in this same equation.
Recent calculations of spherical cooling flows 
(Mathews 1997) have shown that global thermal instabilities
or ``galactic drips'' can occur in isolated ellipticals
if the supernova rate parameter SNu $\lta 0.05$.
Therefore we shall assume
SNu = 0.066 and $E_{sn} = 1$ here although the iron
production may still be somewhat larger than
observed (Mathews 1997).
The time variation for $\alpha_{sn} T_{sn} \propto t^{-1/2}$
is similar to that adopted by Ciotti {\it et al.}
(1991) but less steep.

\newpage

\section{MAGNETIC COOLING FLOWS IN SPHERICAL GEOMETRY}

\subsection{Galactic Model}

We choose a King model for the stellar density 
$$\rho_*(r) = \rho_{o*} [ 1 + (r/r_{c*})^2]^{-3/2}$$
where $r_{c*}$ is the stellar core radius.
Most of the gravitational potential arises from a dark
halo having an approximately isothermal distribution:
$$\rho_{h}(r) = \rho_{oh} [ 1 + (r/r_{ch})^2]^{-1}.$$
Parameters describing the spherical galaxy model listed in
Table 1 have been chosen so that the galaxy lies on
the fundamental plane (see Tsai \& Mathews 1996a
and Brighenti \& Mathews 1996 for
details).
The stellar temperature $T_*(r) = (\mu m_p / 3 k)\sigma_*^2$
can be found by solving the equation of stellar 
hydrodynamics in spherical symmetry (Binney \& Tremaine
1987; Mathews 1988).
For this purpose 
we assume a velocity ellipsoid of the form
$$ { { \langle v_{r}^2 \rangle - \langle v_{tr}^2 \rangle} \over
{ \langle v_{r}^2 \rangle } }
= \left({ r \over r_t}\right)^{q}$$
where $\langle v_{tr}^2 \rangle $ and
$\langle v_{r}^2 \rangle $ are the
transverse and radial stellar
velocity dispersions respectively.
We choose $q = 2$,
but the evolution of the ISM is not sensitive to the value of $q$
over a wide range $1  \lta  q  \lta  3$.

\subsection{Spherical Magnetic Cooling Flows}

In spherical symmetry, the magnetic field can have 
radial and tangential components, ${\bf B} = (B_r,B_t)$
and equation (3.4) becomes
$${ d \log B_r^2 \over d t} = - {4 u \over r}\eqno(4.1)$$
and
$${ d \log B_t^2 \over d t} = - {2 u \over r}
\left( 1 + {\partial  \log u \over \partial \log r}
\right)\eqno(4.2)$$
where $d/dt$ is the Lagrangian or comoving derivative
and $u(r) < 0$ is the radial velocity of the cooling flow.
The (source-free) equation of continuity has a similar 
form:
$${d \log \rho \over dt } = - { u \over r} \left(2 + 
{\partial \log u \over \partial \log r}
\right).\eqno(4.3)$$
Dividing equations (4.1) and (4.2) by (4.3),
we find 
$${d \log B_r^2 \over d \log \rho} 
= {4 \over (2 + \gamma)} \equiv \gamma_r \eqno(4.4)$$
and 
$${d \log B_t^2 \over d \log \rho} 
= {2(1 + \gamma) \over (2 + \gamma)} \equiv \gamma_t 
\eqno(4.5)$$
where
$$\gamma \equiv {\partial \log u \over \partial \log r}.
\eqno(4.6)$$
Therefore each component of the field evolves with the 
local gas density as a power law having an exponent that
depends on the structure the local velocity field
as characterized by the exponent $\gamma$.
In general $\gamma$ can be a function of galactic radius.

For example in a homologous, Hubble-type flow, $u \propto r/t$
and $\rho$ is spatially uniform.
In this case $\gamma = 1$,
$\gamma_r = \gamma_t = 4/3$ so the total magnetic field
$$ B^2 = B_r^2 + B_t^2\eqno(4.7)$$
evolves according to 
$${d B^2 \over d t} = {4 \over 3} {B^2 \over \rho}
{d \rho \over d t}.\eqno(4.8)$$
In typical galactic cooling flows the gas velocity
slowly accelerates toward the galactic center,
$|u| \propto 1 /r^p$ where $0 \lta p \lta 1$.
For this range of velocity fields the radial component
$B_r$ grows faster than $B_t$ toward the 
(high density) center of the cooling flow
(e.g. $B_r^2 \propto \rho^4$, $B_t^2 \sim constant$ for
$p = 0$; $B_r^2 \propto \rho^2$, $B_t^2 \propto \rho$
for $p = 1$).
As a result of this unequal growth,
an initial field random on small scales in non-rotating
ellipticals will
develop radial fields having large scale coherence 
as the flow approaches the galactic core.
Differential shear in slowly rotating ellipticals is expected
to convert disordered fields into coherently toroidal 
configurations (Brighenti \& Mathews 1996). 
In either case, however, reconnection may be expected to 
occur in regions of random field; when such 
disordered fields are stretched
by the global flow, reconnection could still be important since
field lines in adjacent flux tubes will tend to be antiparallel.

The fraction of the field contained in random and coherent
(but antiparallel) 
components in galactic cooling flows can be estimated by
comparing the bulk flow velocity $u$ and the
mean turbulent velocity $v_t$; both velocities are subsonic.
Whenever $B^2$ is less than the equipartion field 
$B_e^2$ and $u \lta v_t$
we expect that the field geometry will be
continuously randomized by local turbulence;
should $u \gta v_t$ the field will be stretched with the
fluid and develop large scale coherence.
The ratio of 
characteristic timescales for turbulence and bulk flow
$t_{turb}/t_{flow} = u/v_t$ can be used to define a parameter
$$\delta(r) = 1 - \exp(-u/v_t)$$
that increases with the development of large scale fields.
Here $\delta \rightarrow 0$ or $\rightarrow 1$
in strong or weak turbulence respectively.
In general we may imagine that the individual field
components have a (source-free) evolution described by
$${d B_r^2 \over dt} = [\delta \gamma_r 
+ (1-\delta) { 4 \over 3}] B_r^2 {1 \over \rho}{d \rho \over dt}$$
and
$${d B_t^2 \over dt} = [\delta \gamma_t 
+ (1-\delta) { 4 \over 3}] B_t^2 {1 \over \rho}{d \rho \over dt}.$$
When continuous randomization obtains,
$\delta \approx 0$ implies 
$B^2 \propto \rho^{4/3}$ and equation (4.8)
holds;
in the following discussion we shall assume this
limit, i.e. the
field is randomized on timescales short compared to the
(quite long) flow time in the interstellar medium.
The validity of this assumption can be estimated by evaluating
$\delta(r)$ from the computed cooling flow solutions.

\subsection{Seed Fields, Turbulent Growth and Reconnection}

While the flux-freezing condition (4.8) describes the evolution 
of a completely disordered field, additional source and sink terms must 
be included to account for the origin, 
turbulent growth and loss of the field:
$${d B^2 \over d t} = {4 \over 3} {B^2 \over \rho}
{d \rho \over d t}
+ {B^2 \over \tau_{turb}} \theta(B^2 - B_e^2)
- {B^2 \over \tau_{rec}}
+ {\alpha \rho_* \over \rho} 
\left({B_*^2 \over \rho_*^{4/3}} \right) \rho^{4/3}
+ \varpi_m {3 \over 2} 8 \pi P \delta(t - t_m). \eqno(4.9)$$

The second term on the right represents the growth of the 
field due to turbulent amplification; the unit step function 
[$\theta(x) = 1$ $x<0$; $\theta(x) = 0$ $x>0$]
ensures turbulent growth only when
$B^2 < B_e^2 = 4 \pi \rho v_t^2$.
This term deals in a schematic fashion with a subject
of considerable complexity.
Normally the turbulent growth of the field would be represented
using the (linear) mean field dynamo equation
(see Beck et al. 1996 and references therein),
however in view of the difficulties described earlier 
in generating equipartition with the mean field equation, 
we prefer the simplified approach
in equation (4.9).
The third term on the right accounts
for the loss of field by reconnection.
The last two terms on the right 
represent original seed fields
from stellar mass loss and galactic events 
at time $t_m$ which 
create non-stellar turbulence in the interstellar gas.
These terms are discussed in more detail in the following.

\subsubsection{Seed Fields}

Aside from the possibility of primordial fields which
are very small in any case (Beck et al. 1996)
the most credible source of magnetic
field in the interstellar medium 
are the fields created by dynamos in stellar interiors 
which are expelled into the ISM 
when red giant stars lose their envelopes.
The relevant magnetic fields are not those observed in stellar
photospheres or chromospheres,
but the mean fields $B_*$ that are likely to exist throughout
(convective) red giant envelopes before ejection from
the star.
Guided by the magnetic flux observed in white dwarfs,
Moss and Shukurov (1996) adopted 
$b_* \equiv B_*^2 / \rho_*^{4/3} 
\approx 2 \times 10^6$
in cgs units as a typical value, 
corresponding to $B_* \approx 10^7$ G 
and $\rho_* \approx 5 \times 10^5 $ gm cm$^{-3}$;
we adopt this same value here.
After the stellar envelope has expanded to the local 
density of interstellar gas $n = \rho/m_p$ conserving magnetic flux,
the seed field will be very small: 
$B_s = b_*^{1/2} \rho^{2/3} \approx 
2 \times 10^{-15} (n/10^{-3})^{2/3}$ G.

An additional source of seed fields may arise
from strong turbulent mixing following a significant merger
of another galaxy with the elliptical at time $t_m$ or after 
some other global environmental disturbance.
Following such an event, we assume that the entire ISM 
becomes strongly turbulent, amplifying any seed fields until
the magnetic energy density is some fraction $\varpi_m$ of
the thermal energy density in the gas.
Such a hypothetical turbulent event is represented 
schematically in equation (4.9)
by a magnetic energy density 
proportional to the local gas pressure,
$B_m^2/8 \pi \approx \varpi_m (3 P/2)$ imposed suddenly at
time $t_m$.

\subsubsection{Turbulent Growth of Field}

The growth of magnetic flux by plasma turbulence 
in the interstellar medium of elliptical galaxies has been
discussed in detail by Moss and Shukurov (1996); we shall
adopt many of their suggestions here.
The two principal internal sources of interstellar turbulence 
are the mixing of gas ejected from 
orbiting stars
and the creation and buoyant evolution of 
(Type Ia) supernova remnants (Mathews 1990).
We assume that the random vortical motion 
associated with these sources 
amplifies the interstellar magnetic field.
The energy density in the field is expected to grow
until it becomes comparable with that in the turbulence
and then saturates at that value.
The equipartition field is defined by
$${ B_e^2  \over 8\pi }
\equiv {1 \over 2} \rho v_t^2 
\equiv \varpi { 3 \over 2} P
= \varpi { 3 \over 2} \rho c_s^2.
\eqno(4.10)$$
The turbulent energy density is written as a fraction
$\varpi(r)$ of the local thermal energy density ($c_s$
is the isothermal sound speed and $v_t$ is an rms speed
characterizing local turbulent motion).
Normally, we expect $\varpi \ll 1$ 
and therefore $B_e^2 /8\pi \ll 3P/2$.
Ideally, a separate dynamical equation should be introduced
to describe the creation, transfer and evolution of turbulent energy
density.
However, we avoid these complications here by assuming
that the turbulent
energy density is generally small ($\varpi < 1$) and
that the principal catalytic role of turbulent energy 
is to convert 
sources of interstellar energy into magnetic energy density.
In addition we shall assume for simplicity 
that the turbulence generated
by stellar ejecta and supernovae is homogeneous
and isotropic; this (somewhat optimistic) assumption is
discussed by Moss and Shukurov (1996).

The turbulent field is produced by a superposition of
kinetic energy created by stellar ejecta
and supernova remnants.
The stellar contribution to the local turbulent
energy density is given by the product of 
the rate of energy production by mass ejection
from evolving (red giant) stars 
$\alpha_* \rho_* ( 3 k T_* / 2 \mu m_p)$ 
(ergs cm$^{-3}$ s$^{-1}$),
the fraction $\epsilon_t$ of this energy that goes into 
turbulence,
and the characteristic
eddy turnover time $\tau = \ell_*/v_t$:
$${1 \over 2} \rho v_{*t}^2 
= \alpha_* \rho_* \epsilon_t 
{3 k T_* \over 2 \mu m_p} { \ell_* \over v_{*t}}
\equiv \varpi_*(r) {3 \over 2} \rho c_s^2.
\eqno(4.11)$$
The characteristic rms turbulent velocity is
$v_{*t} = (3 \varpi_*)^{1/2} c_s$ although the full range
of turbulent velocities must extend up to the local
stellar velocities..
The typical eddy size $\ell_*$ should be comparable to 
the dimension of a stellar envelope
(or planetary nebula) after it has expanded to pressure 
equilibrium (at $T \approx 10^4$K) 
with the local interstellar gas.
For this purpose (following Moss \& Shukurov 1996) we use the 
estimate of this length scale
determined by Mathews (1990),
$\ell_* \approx 4.7 \times 10^{9} \rho^{-1/3}$ cm,
where $\rho$ is the local interstellar density.
We assume that the turbulent mixing time is the eddy turnover
time of the largest eddies, $\ell_* / v_{*t}$;
this is conservative since the turnover rate is faster 
for smaller eddies.
Equation (4.11) can be solved at each galactic radius 
for the stellar equipartition parameter
$$\varpi_*(r) = {1 \over 3 c_s^2}
\left( 3 \ell_* c_*^2 \epsilon_t {\alpha_* \rho_* \over \rho}
\right)^{2/3} \eqno(4.12)$$ 
where $c_*^2 = k T_*/\mu m_p$.
For the elliptical described in Table 1
$T_* \approx 2.95 \times 10^6$ K
(for $10^{0.5} \lta r_{kpc} \lta 10^{1.75}$) and 
$\rho_* \approx 4.35 \times 10^{-22} r_{kpc}^{-3}$ gm cm$^{-3}$.
Field-free cooling flows in this galaxy result in
gas density and temperature distributions given by
$\rho \approx 1.05 \times 10^{-25} r_{kpc}^{-1.8}$
gm cm$^{-3}$ and 
$T \approx 8.91 \times 10^6 r_{kpc}^{-0.25}$ K
at time $t = 15$ Gyr.
These relations result in a stellar equipartition parameter
$\varpi_{*} \approx 1.07 \times 10^{-4}
r_{kpc}^{-0.15} \epsilon_t^{2/3}
(t/t_n)^{-0.87}$ valid for $10^{0.5} \lta r_{kpc} \lta 10^{1.75}$.
The corresponding turbulent velocity induced by stellar ejecta,
$v_{*t} \approx 6.9 \epsilon_t^{1/3} r_{kpc}^{-0.20} (t/t_n)^{-0.43}$
km s$^{-1}$, is indeed very subsonic 
($c_s \approx 385 r_{kpc}^{-0.13}$ km s$^{-1}$).
The turnover time for the largest eddies is
$\tau_{*} = \ell_*/v_{*t} \approx 4.6 \times 10^4 
\epsilon_t^{-1/2} r_{kpc}^{0.8} (t/t_n)^{-0.43}$ years.

In a similar manner Type Ia supernovae feed
energy into the interstellar medium at a rate
$\alpha_* \rho_* (\alpha_{sn} T_{sn} / \alpha_*)
(3 k /2 \mu m_p)$ (ergs cm$^{-3}$ cm$^{-1}$)
and a fraction $\epsilon_t$ of this is assumed to go 
into turbulent motions.
The eddy mixing time is 
$\ell_{sn} / v_{snt}$ where
$v_{snt} = (3 \varpi_{sn})^{1/2} c_s$. 
For supernova-driven turbulence the 
(largest) eddy dimensions 
are comparable to the size of the hot bubble formed
in the interstellar medium following each supernova
event, 
$\ell_{sn} = 3.6 \times 10^{11} \rho^{-1/3}$ cm,
based on the blast wave calculations of 
Mathews (1990).
Combining these expressions, the equipartition
parameter for supernova induced turbulence is
$$ \varpi_{sn}(r) = { 1 \over 3}
\left( \epsilon_t { \rho_* \over \rho } { 3 k \over \mu m_p}
{ \alpha_{sn} T_{sn} \ell_{sn} \over c_s^3 }
\right)^{2/3}.\eqno(4.13)$$
For the bright elliptical shown in Table 1 we find
$\varpi_{sn}(r) \approx 1.11 \times 10^{-3}~
\epsilon_t^{2/3}~
({\rm SNu}/0.01)^{2/3} (t/t_n)^{-1/3} r_{kpc}^{-0.15}$
(valid for $10^{0.5} \lta r_{kpc} \lta 10^{1.75}$),
which is somewhat larger than the stellar contribution
to turbulence for a given $\epsilon_t$.
The corresponding rms turbulent velocity and turbulent
mixing time are 
$v_{snt} \approx 22.2 \epsilon_t^{2/3} r_{kpc}^{-0.20}
(t/t_n)^{-1/6} ({\rm SNu}/0.01)^{1/3}$ km s$^{-1}$
and 
$\tau_{*} = \ell_*/v{*t} \approx 
1.1 \times 10^6 \epsilon_t^{-1/3} r_{kpc}^{0.80}
(t/t_n)^{1/6} ({\rm SNu}/0.01)^{-1/3}$ years.

The combined turbulent mixing time required 
in equation (4.9) is found from 
$$ {1 \over \tau_{turb}}
= {v_{*t} \over \ell_*} + {v_{snt} \over \ell_{sn}}
\eqno(4.14)$$
and the total equipartition field is given by
equation (4.10) with $\varpi(r) = 
\varpi_*(r) + \varpi_{sn}(r)$.
Obviously the introduction of turbulent energy by
supernovae occurs at the 
location of these events, not in a
spatially smooth manner as represented by the 
turbulent source term in equation (4.9).
Nevertheless we shall assume that the turbulence
is driven in a smooth, homogeneous manner.
The applicability of this assumption is discussed 
by Moss \& Shukurov (1996).

In addition to using stellar processes to 
produce vortical turbulence,
Moss and Shukurov (1996) also invoke 
``acoustic turbulence,'' a general field of 
compressional (sound) waves or noise associated with stellar
mass loss and (particularly) supernova explosions.
Although the coupling of this acoustical energy density to 
the usual vortical turbulence is non-linear and
inefficient, Moss and Shukurov claim that 
acoustical wave energy contributes substantially
to the turbulent growth of the magnetic field
and that these waves ultimately steepen into shocks.
However, in the absence of reasonably strong magnetic fields,
a condition that may obtain at early times, viscosity and thermal
conductivity in the hot interstellar medium are 
expected to damp acoustic waves within a wavelength
of their source.
Compression waves should damp before they steepen
into shocks.
For example, damping due to plasma viscosity $\mu$ 
exponentially reduces the flux of
plane parallel sound waves of frequency 
$\omega = 2 \pi c_s / \ell$ 
propagating in the
x-direction by $e^{-x/\ell_{\mu}}$, where 
$$\ell_{\mu} \approx {\rho c_s^3 \over \mu \omega^2}
= 6 \times 10^{15} ~ \ell_{pc}^2
\left( {n \over 10^{-2} {\rm cm}^{-3}}\right) 
\left( {T \over 10^{7} {\rm K} }\right)^{-1/2}
~~{\rm cm}$$
(e.g. Zel'dovich and Raizer 1966).
Damping by thermal conductivity $\kappa$ is comparable.
Evidently acoustic waves generated by either stellar
mass loss or supernova events are strongly 
damped and thermalized
within a single wavelength of their point of origin in 
the interstellar medium. 
Both $\mu$ and $\kappa$ may be reduced
by the presence of a magnetic field, but 
a reduction factor of $\sim 10^{7}$ would be required
for acoustic damping to be negligible,  
requiring the presence of significant magnetic
fields at early times.
In view of this complication we do not consider 
turbulent field generation by the acoustical 
wave field here.

\newpage
\subsubsection{Field Reconnection}

The expected dominance of random magnetic fields
in elliptical galaxies implies that the occurrence
of counter-directed nearby fields should be common.
In such regions thin boundary layers are created where
the finite resistivity converts field energy to
other forms of energy: thermal, bulk kinetic energy,
accelerated particles, etc.
(Soker \& Sarazin 1990; Lesch \& Bender 1990;
Jafelice \& Friaca 1996).
The physics of field reconnection in the hot ISM of ellipticals
galaxies is discussed in some detail by Jafelice and Friaca
where many earlier references are cited.
From dimensional and physical arguments, the flow 
into the reconnection regions 
is expected to occur at some fraction of the
local Alfven velocity $v_a$.
We shall follow the simple 
representation used by these authors for the 
field reconnection time scale in equation (4.9):
$$\tau_{rec} = {\ell_{rec} \over v_a}
= {\ell_{rec} (4 \pi \rho)^{1/2} \over B}.
\eqno(4.15)$$
Here $\ell_{rec}$ is a single parameter that
describes the reconnection efficiency.
In addition we shall assume with these authors that
all of the field energy lost by reconnection is
entirely converted to thermal energy;
so a term $B^2 / 8 \pi \tau_{rec}$ 
must appear in the thermal energy equation (3.3).
The principal parameters related to the turbulent
dynamo are $\epsilon_t$ and $\l_{rec}$ both of which are
very uncertain.

\subsection{Spherical Cooling Flow Equations}

Summarizing the previous results, the equations describing
the time-dependent evolution of spherical cooling
flows containing a self-generated, 
disordered magnetic field are:
$${\partial \rho \over \partial t}
+ {1 \over r^2} {\partial (r^2 \rho u) \over \partial r} 
= \alpha \rho_*,\eqno(4.16)$$
$$\rho {d u \over dt}
= -{ \partial P \over \partial r}
- { 1 \over 12 \pi r} { \partial (r B^2) \over \partial r}
- \rho { \partial \Phi \over \partial r}
- \alpha \rho_* u,\eqno(4.17)$$
$$\rho {d \varepsilon \over d t }
= {P \over \rho} {d \rho \over dt}
- { \rho^2 f L(T) \over m_p^2}
+ \alpha \rho_* \left[ \varepsilon_{ot}
- {P \over \rho}
- \varepsilon + {1 \over 2}
u^2 \right] + {B^2 \over 8 \pi \tau_{rec} }. \eqno(4.18)$$
The source of thermal energy for the gas is determined by 
$$ \varepsilon_{ot} = \varepsilon_o [1 - \epsilon_t
\theta(B^2 - B_e^2)]$$
which allows a fraction $(1 - \epsilon_t)$ of the energy from stellar
processes to be thermalized when $B < B_e$.
When $B > B_e$ all the stellar energy is assumed to 
heat the gas either by direct heating or by eventual 
dissipation of the turbulent component. 
The magnetic field evolves according to 
equation (3.9) which we write as
$${d B^2 \over d t} = {4 \over 3} {B^2 \over \rho}
{d \rho \over d t}
+ \left( {d B^2 \over d t} \right)_{i} 
- {B^2 \over \tau_{rec}}
+ {\alpha \rho_* \over \rho}
\left({B_*^2 \over \rho_*^{4/3}} \right) \rho^{4/3}
+ \varpi_m {3 \over 2} 8 \pi P \delta(t - t_m). \eqno(4.19)$$

We consider two simple representations ($i =1,2$) 
for the turbulent growth of
the magnetic field: 
$$ \left( {d B^2 \over dt} \right)_1 
= {B^2 \over \tau_{turb}} \theta(B^2 - B_e^2)
~~~~~{\rm Case~I}\eqno(4.20a)$$
and
$$ \left( {d B^2 \over dt} \right)_2 
= 8 \pi 
\alpha \rho_* \varepsilon_o \epsilon_t \theta(B^2 - B_e^2).
~~~~~{\rm Case~II} \eqno(4.20b)$$
For Case I the magnetic field can increase only 
on the time scale $\tau_{turb}$ that characterizes turbulent
growth.
The step function $\theta(B^2 - B_e^2)$ guarantees that
the field cannot grow beyond equipartition.
In Case I some part of the power density
$\alpha \rho_* \epsilon_t \varepsilon_o$ is allocated 
to increase the field strength and the rest is assumed to
maintain the local turbulent energy density.

Alternatively, the Case II option allows 
stellar and supernova energy to 
convert directly to either magnetic or thermal energy density,
depending on the value of $\theta(B^2 - B_e^2)$.
The turbulent energy density, which still can be 
regarded as 
an intermediate stage, is not explicitly considered.
The field grows at a rate
$8 \pi \alpha \rho_* \epsilon_t \varepsilon_o$
until it reaches local 
equipartition then growth suddenly saturates.
When equations (4.18) and (4.19) are added in 
the Case II approximation, the total energy is explicitly
conserved either as thermal or magnetic energy.
In the following we consider solutions to both sets
of equations, Cases I and II.

\section{Cooling Flow Models}

We have computed evolving magnetic cooling flows 
in the galaxy described in Table 1
using a Lagrangian code.
The primary parameters that define each flow are
SNu, $\epsilon_t$ and $\ell_{rec}$.
Our cooling flow 
solutions are not sensitive to the value of
the seed field parameter $b_* = B_*^2 / \rho_*^{4/3}$
since for all cases considered 
the exponential growth of field in the turbulent
dynamo is rapid compared to the local
gas flow time ($\sim r/u$) in the ISM.
The parameters $\varpi_m$
and $t_m$ (see eqn. 4.19) can be used to 
simulate the global turbulence expected following 
a galactic merger or some other 
short-lived source of turbulence.
Each calculation is begun at time $t = 1$ Gyr when
it is assumed that 
the galactic winds (driven by Type II supernovae)
have reversed to become cooling flows
(David {\it et al.} 1990; 1991).
Interstellar turbulence is turned on at a 
later time ($t = 2$ Gyr)
to avoid numerical interference that may occur if the
gas and the magnetic field are initiated at the same time.
The calculations are stopped at 15 Gyr 
which we regard as the present time.
Our elliptical galaxy is assumed to be perfectly isolated 
apart from the environmental influences represented by the 
parameters $\varpi_m$ and $t_m$.

In Figure 1 we compare several 
cooling flow models at time $t = 15$ Gyr.
The non-magnetic flow shown in Figure 1a-c resembles 
spherical cooling flows 
that have been discussed in detail elsewhere 
(e.g. Brighenti \& Mathews 1996; Mathews 1997).
The parameters used for this non-magnetic cooling flow are:
$\epsilon_t = 1$ and SNu $= 0.066$.
Some numerical irregularity is visible in Figures 1a and 1b for
$\log r \lta 2.5$ because of the smaller number of computational
zones in the galactic core.
Most of this numerical noise arises when central
zones are removed as they cool below $T = 10^4$ K.
We recognize that the x-ray surface brightness 
$\Sigma_x(r)$ corresponding to 
our model in Figure 1a-c decreases 
somewhat more steeply with galactic radius
than that typically observed 
(Trinchieri, Fabbiano, \& Canizares 1986).
The traditional correction for this discrepancy has been to
remove 
gas from the flow, assuming that stars or some other invisible 
objects are formed from the mass that ``drops out''
(e.g. Stewart et al. 1987; White \& Sarazin 1987;
Thomas et al. 1987).
Our models would adjust in a similar manner if mass were removed.
We have not included drop out in our calculations for several
reasons: (i) we wish to illustrate the possible dynamic influence of
magnetic forces without additional 
uncertain complications, (ii) in those 
(perhaps unlikely) situations when the magnetic field is strong, 
its influence on the solution (and $\Sigma_x(r)$) is very similar
to mass drop-out, and (iii) thermal instabilities, often
invoked to justify mass drop out, are generally arrested at 
a very early stage in the presence of modest magnetic fields.

The corresponding Case I magnetic cooling flow 
evolution with parameters $\epsilon_t = 0.5$ 
SNu $= 0.066$, $\ell_{rec} = 3 \times 10^{30}$ cm, and
$\varpi_m = 0$
is illustrated by the solid lines in Figure 1d-f.
Our choice of a large
$\ell_{rec}$ effectively shuts down the reconnection process,
maximizing the computed field strength.
Even though half of the energy from stellar sources is
available to drive the turbulent dynamo,
the gas density, temperature and pressure are all 
essentially unchanged from 
the field-free cooling flow (Figure 1a-c).
The magnetic pressure $B^2/8 \pi$ (long dashed line in Fig. 1f) 
is never more than a few percent of the gas pressure
although it is somewhat larger at small galactic radii.
But  the field strength indicated in Figure 1f 
(long-dashed curve) is not small,
reaching 1 $\mu$G at $r = 7$ kpc and 10 $\mu$G at 1 kpc.
Of course these fields are upper limits because we assume 
that the dynamo is efficient and our choice of 
large $\ell_{rec}$
effectively deactivates field reconnection.
We find that the results in Figure 1f are essentially
unchanged as long as
$\ell_{rec} \gta 10^{25}$ cm.
The magnetic field throughout the galaxy slowly decreases 
with $\ell_{rec}$; 
in Figure 1f we show that the field is reduced only 
by a factor of $\sim 2.5$ (short-dashed line) when
$\ell_{rec} = 10^{23}$ cm.
Unfortunately the physical significance of this is
unlear.
The reconnection process, if it is important,
is expected to occur throughout the flow in very thin boundary
layers separating regions of adjacent antiparallel fields.
Such a physically complex process cannot be adequately modeled with
the single parameter $\ell_{rec}$. 
Also illustrated in Figure 1f 
(dotted line) is the magnetic pressure 
at $t = 15$ Gyr that results when 
$\epsilon_t$ is increased to $0.8$
with the other parameters unchanged
(SNu $= 0.066$, $\ell_{rec} = 3 \times 10^{30}$ cm, and
$\varpi_m = 0$).
The field is slightly larger in this case because of the somewhat
greater turbulent energy that existed early in the cooling flow
evolution.

We have also computed cooling flow solutions with the same 
parameters but using the Case II equations.
We find that 
the results are essentially identical to 
those of the Case I solutions.
As explained below, 
the insensitivity to the specific representation 
(Equation 4.20a or b) for the 
turbulent growth of the field
arises because for either Case the field grows to 
equipartition early in the calculation 
in a time short compared to the local flow 
time ($\sim r/u$).

Of particular interest for 
understanding the amplification of disordered
fields in galactic cooling flows are the radial dependences
of the thermal, magnetic and turbulent energy densities.
Using the approximate power law variations discussed in
\S 4.3.2, the thermal energy density should vary as
$\epsilon_{therm} \propto \rho T \propto r^{-2.05}$.
For turbulence driven by either stars or Type Ia supernovae,
the turbulent energy density varies as 
$\epsilon_{turb} = 
\rho v_t^2/2 \propto \rho \varpi c_s^2 \propto r^{-2.20}$
and the equipartition field must vary in the same manner,
$B_e^2 \propto r^{-2.20}$.
However, the field compression that accompanies the cooling 
flow causes a slightly steeper variation,
$\epsilon_{mag} \propto B^2 \propto \rho^{4/3} \propto r^{-2.40}$.

Two important conclusions follow from these spatial
dependencies: (1) the increase in the magnetic field strength 
toward the galactic center can be dominated by compression
in the cooling flow, not local turbulence, and, if so, (2) the
magnitude of the interstellar field depends entirely on 
the most distant radius in the galactic ISM where 
the turbulent dynamo process amplified stellar seed fields 
at some previous time. 

In Figure 2a we plot the spatial variation of the thermal,
magnetic and turbulent energy densities for the 
standard magnetic cooling flow 
solution (shown in Figure 1d-f with solid lines) 
at times $t = 12$ and $15$ Gyr; these contours are essentially
identical for Case I and Case II equations.
The slopes of the various energy
densities (in the linear parts of the log-log contours)
are in excellent agreement with the approximate variations
estimated above.
At either 12 or 15 Gyrs 
$\epsilon_{mag}$ is almost 10 times larger than 
$\epsilon_{turb}$ throughout most of the galactic volume.
As expected from the previous estimate, 
the local turbulent energy density
plays no explicit role in field amplification since $B^2 > B_e^2$.
This excess of the field above the local equipartition value 
occurs because the field at $t = 15$ Gyr is a relic of a 
historical field which developed when the turbulence was larger.
At an early time in the evolution of this cooling flow
equipartition ($\epsilon_{mag} \approx \epsilon_{turb}$) was reached 
in time $\sim \tau_{turb}$ throughout 
the galaxy, but as the calculation
proceeded the turbulent energy decreased 
($\epsilon_{turb} \propto \rho v_{tsn}^2 \propto t^{-1/3}$
as expected from the scaling relations 
in \S4.3.2 when $v_{tsn}^2 \gta v_{t*}^2$).
The super-equipartition 
field is maintained and amplified by compression in the cooling flow.
Since $\epsilon_{turb}/\epsilon_{mag}$ increases
with radius, the current galactic field is most sensitive to 
past turbulent
conditions in the outermost parts of the galactic flow.

In this regard we note 
the small decrease in $\epsilon_{mag}$ in Figure 2a 
that occurs near the outer boundary $r_t$ of the galaxy 
between 12 and 15 Gyrs.
In this part of the flow the gas density drops slightly during
this time interval due to the (losing) competition between 
the decreasing rate that gas is introduced by evolving stars and the
depletion of gas by global inflow toward the galactic core.
As $\rho$ decreases,
the magnetic field must also decrease ($B^2 \propto \rho^{4/3}$).
Some additional loss of local field near $r_t$ may occur 
because the local field is being diluted by new gas that
only contains the very small seed field yet 
turbulent amplification is not
possible until $B^2$ drops below the equipartition value.
Figure 2b shows 
how the frozen-in field has been advected inward toward the
galactic center, creating and maintaining the expected slope $B^2 \propto
\rho^{4/3}$ throughout the interstellar medium.
This also illustrates how the field throughout the 
X-ray bright central
parts of the ISM depends critically on the historical turbulent
growth in very distant regions.
In view of the relationship between the global field
and past turbulence near the outer edge of the galaxy,
it might be thought that the growth of the field would
be sensitive to the outer boundary conditions in the flow,
but this is not the case.
For example we repeated the standard calculation 
shown in Figure 1d-f (solid lines) with 
an open boundary condition at $r = r_t$ which allows gas
to flow out beyond the galaxy into the surrounding vacuum.
After $t = 15$ Gyrs 
the flow variables and the field strength distribution 
within the galaxy were essentially identical
to those shown in Figure 1d-f.
Evidently the gas density and pressure near $r \lta r_t$ 
is maintained by local stellar mass loss, not by 
the small amount 
of gas that flows out of the galaxy when an open boundary
is allowed.

Finally, we have performed some calculations with non-zero
$\varpi_m$ to simulate an event, such as a galactic merger,
that introduces global turbulence throughout the ISM at
some time $t_m$ in the past. 
In Figure 1g-i we illustrate the magnetic cooling flow
at time $t = 15$ Gyr following 
a disturbance in the the interstellar 
gas that occurred at $t_m = 11$ Gyr when the 
turbulent energy density was raised to one fourth
of the thermal energy density ($\varpi_m = 0.25$)
throughout the galaxy.
As a result of this event, 
the magnetic energy density at 15 Gyr 
(Fig. 1i) is $\sim 100$ times
larger than it would have been otherwise (Fig. 1f).
This illustrates that present day magnetic fields in elliptical
galaxies can be significantly increased by past events
that stirred up the outer parts of the interstellar gas
and which expended energies that are small compared to 
the total binding energy of the interstellar gas.
For the cooling flow in Figure 1g-i the magnetic energy density
$\varpi_m 3P/2 = 3P/8$ 
is imposed impulsively at time $t_m$ and subsequent
growth of $B^2/8 \pi$ is due to compression in the global
flow.
By time $t = 15$ Gyr Figure 1i shows that magnetic pressure
dominates the flow dynamics within about 5 kpc of the galactic 
center while the gas density remains monotonic, implying that 
the magnetic core does not become buoyant.
However, the magnetic stresses are sufficient to slow down the 
local flow in $r \lta 5$ kpc allowing an accumulation of gas
near this radius where some 
radiative cooling and mass deposition occurs.
This is reminisent of the ``galactic drip'' phenomenon 
in which gas in non-magnetic 
cooling flows can cool far from the galactic 
center (Mathews 1997), but the increase in gas density 
and thermal instability shown in Figure 1i has a purely 
magnetic origin. 
When $\varpi_m$ is lowered to 0.15,
$B^2/8 \pi \approx P$ occurs at a smaller radius, 
$r \approx 1$ kpc, and radiative cooling beyond the 
galactic core no longer occurs.

Some caution must be exercised in interpreting the 
non-stellar turbulent growth imposed at $t_m$ 
in terms of a galactic merger.
For galactic mergers in which the galactic mass 
is increased by a relatively small factor, $\sim 10$ percent, 
Mathews and Brighenti (1997) have shown that enough energy can 
be released by Type II supernovae from newly created stars 
to remove most or all of the 
preexisting galactic interstellar gas. 
Such an energy release obviously depends on the assumed 
initial mass function for the newly formed stars.
We do not consider these complications here.

\section{DISCUSSION AND CONCLUSIONS}

In this study of the evolution of interstellar magnetic
fields in elliptical galaxies
we have shown that the observed fields can be understood
as a natural result of internal evolutionary processes:
stellar seed fields are first amplified by 
a turbulent dynamo then by 
compression in the galactic cooling flow.
We have assumed that stellar dynamos occur in virtually
all stars and that these stars 
supply tiny seed fields to the ISM as they undergo mass loss 
during normal red giant evolution.
Our optimism concerning the efficacy of the turbulent
dynamo mechanism and the relative inefficacy of field reconnection 
have been essential in obtaining 
central field strengths comparable to those observed, 
$\sim 1 - 10$ $\mu$G.

Our models extend a modified version of the
turbulent dynamo theory proposed by
Moss and Shukurov (1996) to include the important effects
of compressional 
field amplification in the global galactic cooling flow
and the evolutionary relationship between currently observed fields
and those created by galactic turbulence many Gyrs ago.
As the field advects from the outer regions of the galaxy
to the galactic core the field compresses by $\gta 10^3$ 
and the magnetic energy density increases by $\gta 10^6$.
In our models 
the magnetic energy density exceeds the
local turbulent energy density ($B^2 > B_e^2$) 
throughout most of the interstellar gas so 
local turbulence plays little or no role in field amplification.
When $B^2 > B_e^2$ the field may quench 
some of the turbulent activity although we have not allowed for 
this feedback in our models; it is also unclear whether the 
random character of the field 
(on scales of $\ell_*$ or $\ell_{sn}$) 
can be fully maintained in this limit.
For most of the magnetic cooling flows described here 
we have assumed that half of the stellar energy goes 
(at least initially) into
turbulence ($\epsilon_t = 0.5$).
Although the ratio $\epsilon_t$ of turbulent energy to immediately 
thermalized energy resulting from stellar mass loss
processes is uncertain,
our results suggest that the resulting field strength 
does not depend strongly on $\epsilon_t$.

How sensitive are our results, particularly the 
condition $B^2 > B_e^2$, to the specific 
galaxy model we have adopted, a simple King stellar distribution
with an isothermal dark halo?
To address this question it is useful to 
express the radial variation of the magnetic and turbulent
energy densities in terms of the gas and stellar densities.
For either stellar or supernova 
turbulence we find $\epsilon_{turb} = \rho v_t^2 /2
\propto \rho^{1/9} \rho_*^{2/3}$ 
and $\epsilon_{mag} \propto B^2 \propto \rho^{4/3}$.
Therefore the ratio of these energy densities
$\epsilon_{mag}/\epsilon_{turb} 
\propto \rho^{11/9} \rho_*^{-2/3}$.
Clearly the radial variation of the gas density 
depends on both the stellar and dark matter 
density profiles, so the interpretation of this
last relation is complex.
To make further progress, suppose that we adopt
the point of view that the X-ray and optical
surface brightness profiles are the same,
i.e. $\Sigma_* \propto \Sigma_x$.
Canizares, Fabbiano, \& Trinchieri (1987) found
this to be the case for three bright ellipticals
in Virgo, but more recent 
ROSAT observations show some deviations 
from exact proportionality.
However, if $\Sigma_* \propto \Sigma_x$ then
$\rho^2 \propto \rho_*$ and 
$\epsilon_{mag}/\epsilon_{turb}
\propto \rho^{-1/18}$ so that $\epsilon_{mag}$
{\it increases} very slowly relative to $\epsilon_{turb}$
with increasing galactic radius, opposite to the 
variation shown in Figure 2a.
We may expect, therefore, that the ratio
$\epsilon_{mag}/\epsilon_{turb}$ could either
slowly increase or decrease with galactic radius,
depending on the details of the galactic model.
However, regardless of these details it is clear that 
the interstellar magnetic field can
increase toward the galactic center {\it no slower}
than $B^2 \propto \rho^{4/3}$ as shown in Figure 2b. 
Finally we note that
in our models the $\epsilon_{mag}/\epsilon_{turb}$ 
increases with the cooling inflow most strongly in the outermost
parts of the galaxy where the power law scaling breaks
down and where the observational uncertainty is greatest.

If interstellar fields in elliptical galaxies have their 
origins in the seed fields of stellar ejecta, the physical 
scale $d$ of regions of Faraday coherence may be smaller than
previously supposed (Burn 1966; Garrington \& Conway 1991). 
Regions in the hot interstellar gas having masses comparable 
to those of stellar envelopes, $0.4$ $M_{\odot}$, have 
diameters of only 
$d \approx 30 (n/10 \times 10^{-3})^{-1/3}$ pc. 
However, Faraday-coherent 
regions of parallel field lines will be even smaller 
because of the violent instabilities that accompany stellar 
mass loss.
Since the observed Faraday depolarization parameter is 
proportional to 
$\langle n B_{\parallel} \rangle (d r_t)^{1/2}$, the
field implied by any observation will be larger as $d$ 
is reduced.
This difficulty is not alleviated by large scale fields that
can develop in (non-Hubble) spherical 
or differentially rotating non-spherical cooling 
flows, since such an ``aligned'' field will still be 
counter-directed on small mass scales.
The scale of field coherence $d$ can increase if 
reconnection is efficient,
but this would also decrease the overall magnetic energy density.
Clearly, more attention must be given to the likelihood of
field reconnection in cooling flow environments.

Our main conclusions are:

\noindent
(1) Self-generated interstellar magnetic fields
in elliptical galaxies can be comparable to those 
suggested in Faraday depolarization
studies provided
(i) the turbulent dynamo process can amplify the mean
field to equipartition with the turbulent energy density 
and (ii) field reconnection is not dominant throughout
most of the ISM.

\noindent
(2) Using our galactic model,
a significant additional amplification of the overall field 
occurs by compression in the galactic cooling flow; the 
magnetic energy density is 
increased by factors of $\gta 10^6$.

\noindent
(3) Throughout most of the galactic volume
the field energy density can exceed local equipartition 
with the turbulent energy density.

\noindent
(4) The exact value of the field in ellipticals is sensitive
to interstellar turbulence at large distances 
from the galactic center
and at distant times in the past.
For this reason we may expect some variability in field strengths 
among elliptical galaxies, 
depending on the environmental history of the outer galaxy.

\noindent
(5) In our spherical cooling flow 
models the magnetic energy density or pressure 
in self-generated cooling flows ($\varpi_m = 0$) 
never dominates the galactic gas pressure 
(nor becomes buoyant) so we do not 
expect the X-ray images to be strongly influenced by the field.
In more realistic rotating cooling flows, however, 
the magnetic pressure resulting from differential 
rotation may decrease the X-ray brightness near the 
cooling flow disk (Brighenti \& Mathews 1996).

\noindent
(6) If the interaction of the elliptical 
with ambient gas or nearby 
galaxies stimulates strong turbulence in the galactic ISM, 
the resulting field strength may be greatly
enhanced many Gys afterward.

\vskip.4in
\noindent
We are pleased to acknowledge comments from Niel Turner 
and support from 
a UCSC Faculty
Research Grant and NASA grant NAG 5-3060.

\newpage

\noindent
\centerline {\bf References}\\

\noindent
Balbus, S. A. 1991, ApJ, 372, 25\\
Beck, R., Brandenburg, A, Moss, D., Shukurov, A.,
\& Sokoloff, D. 1996, {\it Ann. Rev. Astron. \& Astrophys.}, 
34, 155\\
Binney, J. \& Tremaine, S. 1987, Galactic Dynamics 
(Princeton: Princeton Univ. Press), 195\\
Brighenti, F. \& Mathews, W. G. 1996, ApJ, 470, 747\\
Burn, B. J. 1966, MNRAS, 133, 67\\
Canizares, C., Fabbiano, G., \& Trinchieri, G. 1987, ApJ, 312, 503\\
Cappelaro, E., Turatto, M., Benetti, S., Tsvetkov D.Yu.,
Bartunov, O. S., \& Makarova I. N., 1993, A\&A, 273, 383\\
Christodoulou, D. M. \& Sarazin, C. L. 1996, ApJ, 463, 80 \\
Clarke, D. A., Burns, J, O., \& Norman, M. L. 1992,
ApJ, 395, 444\\
David L. P., Forman, W., \& Jones, C. 1990, ApJ, 359, 29\\
David L. P., Forman, W., \& Jones, C. 1990, ApJ, 369, 121\\
Donnelly, R. H., Faber, S. M., \& O'Connell, R. M. 1990, ApJ, 354,
52\\
Field, G. B. 1995, in {\it The Physics of the Interstellar 
Medium}, ASP Conf. Series vol. 80, 1\\
Garrington, S. T., Conway, R. G. \& Leahy, J. P., 1991,
MNRAS, 250, 171\\
Garrington, S. T. \& Conway, R. G., 1991, MNRAS, 250, 198\\
Garrington, S. T., Leahy, J. P., Conway, R. G. \& Laing, R. A.,
1988, Nature, 331, 147\\
Greenfield, P. E., Roberts, D. H., \& Burke, 1984, ApJ, 
293, 370\\
Jafelice, L. C. \& Friaca, A. C. S. 1996, MNRAS, 280, 438 \\
Kim, Dong-Woo, \& Fabbiano, G. 1995, ApJ, 441, 182\\
Kulsrud, R. M. \& Anderson, S. W. 1992, ApJ, 396, 606\\
Laing, R. A., 1988, Nature, 331, 149\\
Lesch, H. \& Bender, R. 1990, A\&A 233, 417\\
Lesch, H. \& Chiba, M. 1995, A\&A, 297, 305\\
Mathews, W. G. 1988, AJ, 95, 1047\\
Mathews, W. G. 1990, ApJ, 354, 468 \\
Mathews, W. G. 1997, AJ, (in press)\\
Mathews, W. G. \& Brighenti, F. 1997, in {\it The Nature of
Elliptical Galaxies, Proc. of the Second Stromlo Symposium},
Eds. M. Arnaboldi, G. S. Da Costa \& P Saha, 
(San Francisco: ASP), A.S.P. Conf. Ser. (in press)\\
Moss, D. \& Shukurov, A. 1996, MNRAS, 279, 229 \\
Owen, F.N., Eilek, J. A. \& Keel, W. C., 1990, ApJ, 362, 449\\
Rees, M. 1987, QJRAS, 28, 197\\
Ruzmaikin, A. A., Shukurov, A. M. \& Sokoloff, D. D. 1988,
{\it Magnetic Fields of Galaxies}, (Dordrecht: Kluwer)\\
Soker, N. \& Sarazin, C. L. 1990, ApJ, 348, 73 \\
Stewart, G. C., Canizares, C. R., Fabian, A. C., \& Nulsen, P. E. J.
1987, ApJ, 278, 536\\
Strom, R. G. \& J\"{a}gers, 1988, A\&A, 194, 79\\
Thomas, P. A., Fabian, A. C., \& Nulsen, P. E. J. 1987, MNRAS, 228,
973\\
Trinchieri, G., Fabbiano, G, \& Canizares, C. R. 1986, ApJ, 310, 637\\
Tsai, J. C. \& Mathews, W. G. 1996a, ApJ, 448, 84 \\
Tsai, J. C. \& Mathews, W. G. 1996b, ApJ, 468, 571 \\
Vainshtein, S. I. \& Cattaneo, F. 1992, ApJ, 393, 165\\
van den Bergh, S., McClure, R. D., \& Evans, R. 1987, ApJ, 323, 44\\
van Dokkum, P. G. \& Franx, M. 1995, AJ, 110, 2027\\
White, R. E. III, \& Sarazin, C. L. 1987, ApJ, 318, 612\\
Zel'dovich, Ya. B., \& Raizer, Yu. P. 1966, Physics of shock waves
and High-Temperature Hydrodynamic Phenomena, (New York: Academic
Press), 74


\begin{planotable}{l l}
\tablewidth{4.0in}
\tablenum{1}
\tablecaption{GALACTIC PARAMETERS FOR ELLIPTICAL GALAXY}
\tablehead{
\colhead{Parameter} &
\colhead{Value}
}

\startdata

$r_{c*}$        &       311.59 pc\cr
$r_e$\tablenotemark{a}        &       5.088 kpc\cr
$r_{ch}$         &       6.22 kpc \cr
$r_t$           &       113.1 kpc\cr
$\rho_{*o}$     &       $1.438 \times 10^{-20}$~ gm cm$^{-3}$\cr
$\rho_{ho}$  &       $5.46 \times 10^{-24}$~ gm cm$^{-3}$\cr
$M_{*t}$        &       $4.52 \times 10^{11}$ $M_{\odot}$\cr
$M_{ht}$        &       9 $M_{*t}$\cr
$L_B$           &       $4.95 \times 10^{10} L_{B\odot}$\cr
$M_{*t}/L_B$    &       9.14\cr
$\sigma_*$\tablenotemark{b}   &       351 km s$^{-1}$\cr
$\log L_x$\tablenotemark{c}   &       40.0\cr

\tablenotetext{a}{Effective radius.
\vskip.2in}

\tablenotetext{b}{Characteristic velocity dispersion in stellar core,
$\sigma_* = ( 4 \pi G \rho_{*o} r_{c*}^2 /9 )^{1/2}$.
\vskip.2in}

\tablenotetext{c}{For 0.5 - 4.5 keV based on correlation of
Donnelly, Faber \& O'Connell (1990);
more recent $L_x$ values are higher (Kim \& Fabbiano 1996).}

\end{planotable}

\newpage

\noindent
\centerline{\bf Figure Captions}\\

\noindent
{\bf Figure 1:} Three vertical arrays of figures show cooling 
flow conditions in the interstellar gas 
after evolving to 
$t = 15$ Gyr.
Solid lines in each plot refer to the gas density,
temperature or pressure corresponding to these parameters:
(a-c), nonmagnetic cooling flow
with $\epsilon_t = 0.5$ and SNu $= 0.066$; 
(d-f), magnetic cooling flow with 
$\epsilon_t = 0.5$, SNu $= 0.066$,
$\ell_{rec} = 3 \times 10^{30}$ and $\varpi_m = 0$;
(g-i), magnetic cooling flow with
$\epsilon_t = 0.5$, SNu $= 0.066$,
$\ell_{rec} = 3 \times 10^{30}$, $\varpi_m = 0.25$
and $t_m = 11$ Gyr.
Long-dashed lines in the density plots show the variation of 
the (unnormalized) stellar density.
Long-dashed lines in the temperature plots show the variation of
stellar temperature.
Long-dashed lines in the pressure plots show the magnetic
pressure.
Plot (f) also shows the magnetic pressure with $\epsilon_t = 0.8$
(dotted line) and with some field reconnection 
$\ell_{rec} = 10^{23}$ (short-dashed line).
See text for further details.

\noindent
{\bf Figure 2:} (a) Variation of energy densities with galactic
radius at $t = 12$ Gyr ({\it thick lines}) and $t = 15$ Gyr 
({\it thin lines}): $\epsilon_{therm}$ ({\it solid lines}),
$\epsilon_{mag}$ ({\it dotted lines}), and 
$\epsilon_{turb}$ ({\it dash-dotted lines}).
(b) Variation of magnetic energy density with gas density 
at $t = 12$ Gyr ({\it dashed line}) and $t = 15$ Gyr ({\it solid line}).
The flow parameters are: $\epsilon_t = 0.5$, SNu $= 0.066$,
$\ell_{rec} = 3 \times 10^{30}$ and $\varpi_m = 0$.

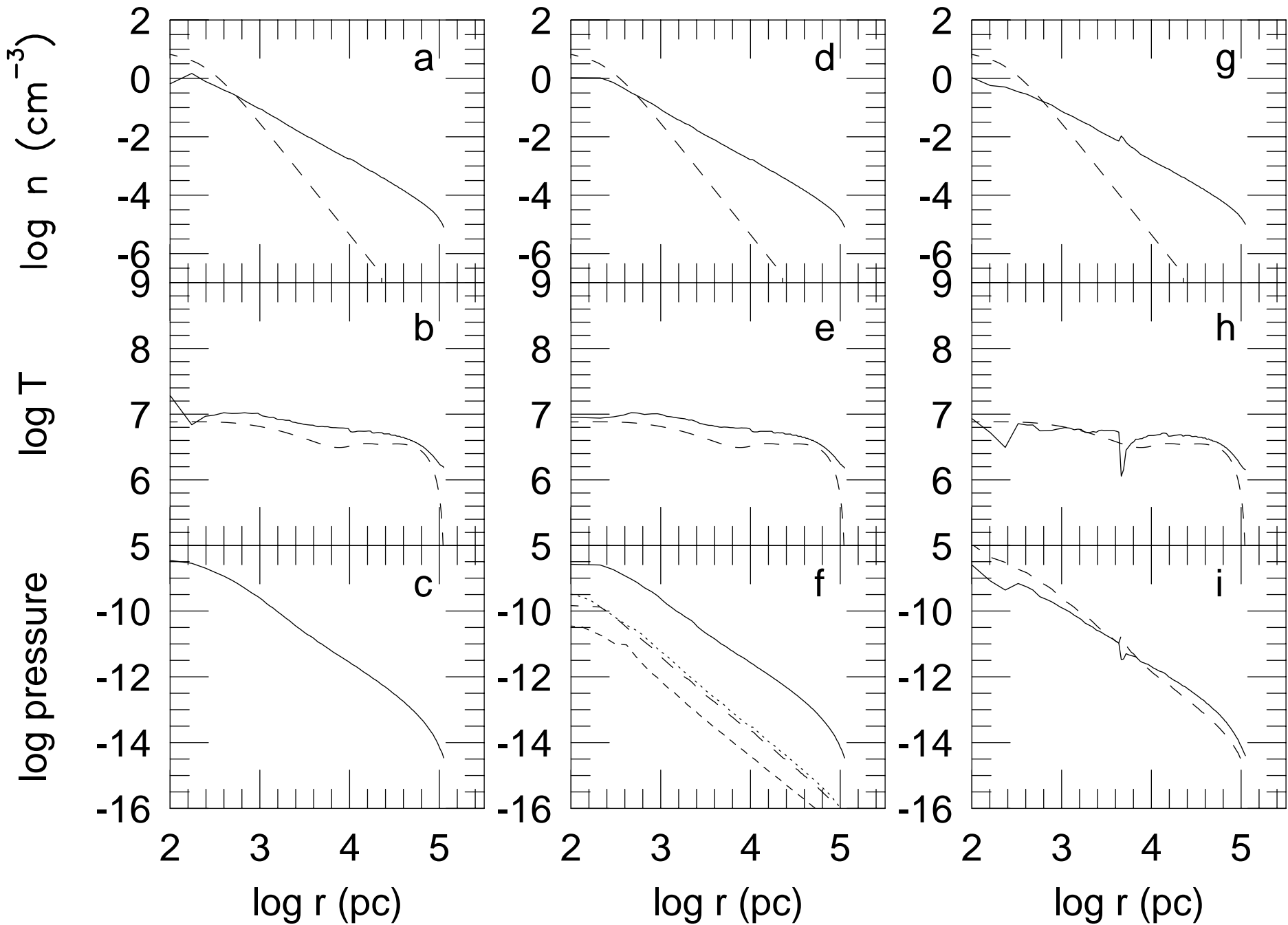
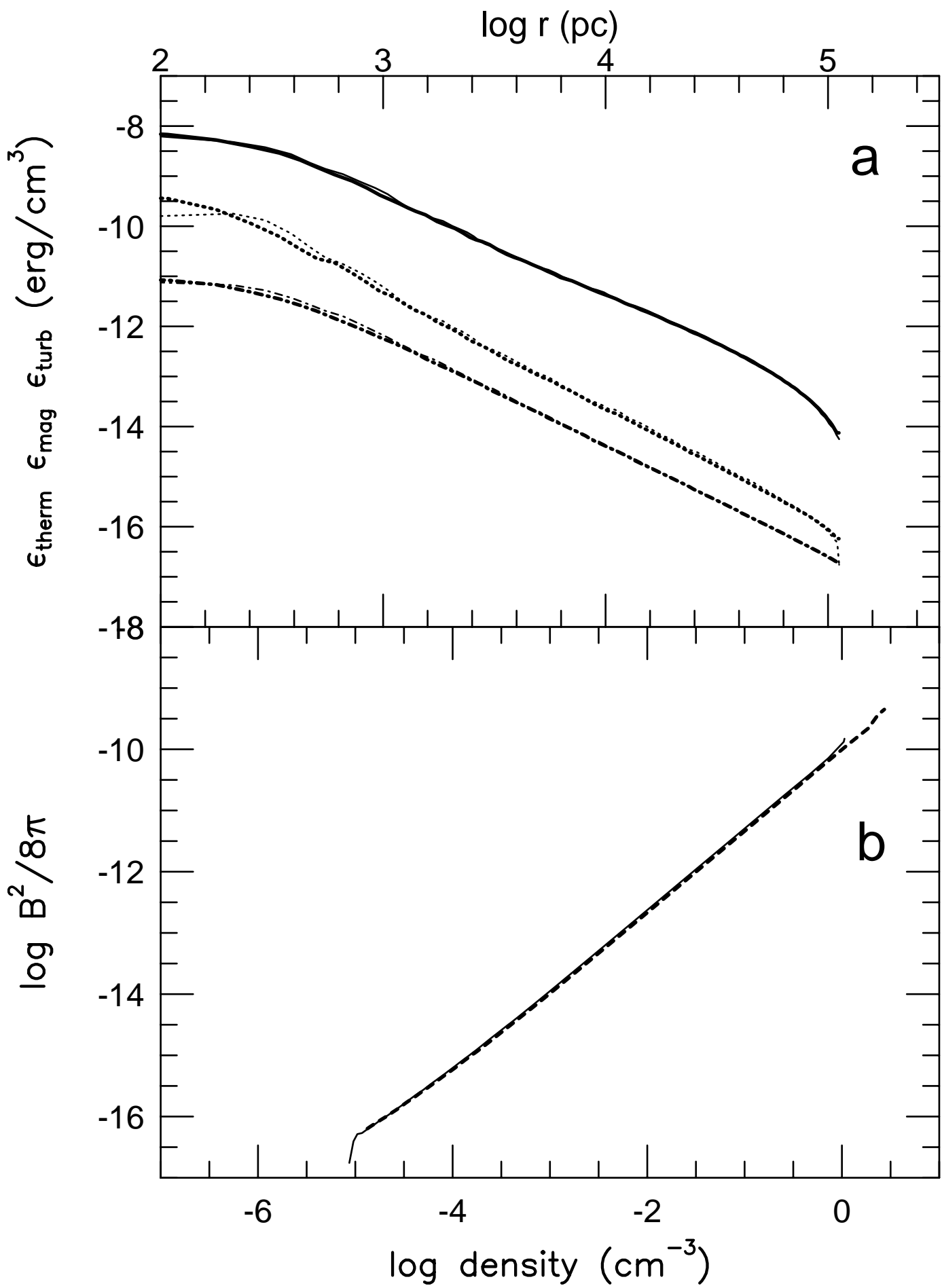
\end{document}